\pdfoutput=1
\documentclass[11pt,a4paper]{article}   
\usepackage[utf8]{inputenc}
\usepackage{fourier} 
\usepackage{fullpage}
\usepackage{enumerate}
\usepackage{wrapfig} 
\usepackage{mathrsfs}
\usepackage[usenames,dvipsnames]{xcolor}
\usepackage[colorlinks=true,linkcolor=Blue,citecolor=Green,urlcolor=Black]{hyperref}
\usepackage{amsmath,amsthm,amssymb}
\usepackage{mathtools}
\usepackage[boxed, ruled]{algorithm2e}
\usepackage{framed}

\usepackage[autostyle]{csquotes}
\usepackage[
backend=biber,
style=alphabetic,
maxbibnames=99,
maxalphanames=99,
maxnames=99,
sorting=nyt,
firstinits=true,
url=false,
doi=false,
eprint=true,
isbn=false
]{biblatex}
\usepackage{utf8math}
\usepackage{todonotes}

\newtheorem*{theorem*}{Theorem}

\newtheorem{theorem}{Theorem}[section]
\newtheorem{lemma}[theorem]{Lemma}

\theoremstyle{definition}

\theoremstyle{remark}
\newtheorem{remark}[theorem]{Remark}

\newcommand{\norm}[1]{\left\|#1\right\|}
\newcommand{\cvp}[1][]{\ensuremath{\mathrm{CVP}_{#1}}}
\newcommand{\svp}[1][]{\ensuremath{\mathrm{SVP}_{#1}}}
\newcommand{\acvp}[2][(1+\epsilon)]{\ensuremath{#1}-\cvp[#2]}

\renewcommand{\epsilon}{\varepsilon}

\newcommand{\vs}{\mathbf{s}}
\newcommand{\vv}{\mathbf{v}}
\newcommand{\vy}{\mathbf{y}}
\newcommand{\vt}{\mathbf{t}}
\newcommand{\vw}{\mathbf{w}}
\newcommand{\vx}{\mathbf{x}}

\newcommand{\vb}{\mathbf{b}}

\newcommand{\wt}[1]{\widetilde{#1}}

\DeclareMathOperator{\listred}{\mathtt{ListRed}}

\def\R{\mathbb{R}}
\def\Q{\mathbb{Q}}

\def\Z{\mathbb{Z}}

\def\N{\mathbb{N}}

\DeclareMathOperator{\vol}{vol}

\DeclareMathOperator{\poly}{poly}

\DeclareMathOperator{\binentropy}{H}

\author{Friedrich Eisenbrand \footnote{The  author acknowledges support from the Swiss National Science Foundation (SNSF) within the project \emph{Lattice Algorithms and Integer Programming (Nr. 185030)}.}\\
 EPFL\\
 Switzerland\\
 {\small \texttt{friedrich.eisenbrand@epfl.ch}}
\and
 Moritz Venzin \\
 EPFL\\
 Switzerland\\
 {\small \texttt{moritz.venzin@epfl.ch}}
}
\date{\today}

\title{Approximate \cvp[p]  in time $2^{0.802 \, n}$} 

\begin{document} 
\maketitle
\begin{abstract}
  \noindent 
    We show that a constant factor approximation of the shortest and closest lattice vector problem w.r.t. any $\ell_p$-norm 
  can be computed in time $2^{(0.802 +ε)\, n}$. This matches the currently fastest constant factor approximation algorithm for the shortest vector problem w.r.t. $\ell_2$. To obtain our result, we combine the latter algorithm w.r.t. $\ell_2$ with geometric insights related to coverings. 

\end{abstract}

\section{Introduction}

The \emph{shortest vector problem} (SVP) and the \emph{closest vector
  problem} (CVP) are important algorithmic problems in the geometry of
numbers. Given a rational lattice
\begin{displaymath}
  ℒ(B) = \{B\vx ： \vx \in \Z^n\}
\end{displaymath}
with  $ B \in \Q^{n \times n}$ and a target vector $\vt ∈ ℚ^n$ the closest vector problem asks for lattice vector $\vv ∈ ℒ(B)$  minimizing $\norm{\vt-\vv}$. 
 The \emph{shortest vector problem} asks for a \emph{nonzero} lattice vector $\vv ∈ ℒ(B)$ of minimal norm. 
When using the $\ell_p$ norms for $1 \leq p \leq \infty$, we denote the problems by $\text{SVP}_p$ resp. $\text{CVP}_p$. 

Much attention has been devoted to the hardness of approximating \svp[] and \cvp[]. In a long sequence of papers, including~\cite{vanEmdeBoas81,svp_hard_ajtai,micciancio2001shortest, Arora:1995:PCP:220989,DBLP:journals/combinatorica/DinurKRS03,svp_hard_khot,svp_hard_regevhaviv} it has been shown that \svp[] and \cvp[] are hard to approximate to within almost polynomial factors under reasonable complexity assumptions. The best polynomial-time approximation algorithms have exponential approximation factors~\cite{LLL,schnorr1987hierarchy,DBLP:conf/stoc/AjtaiKS01}.

The first algorithm to solve \cvp[] for any norm that has exponential running time in the dimension only was given by Lenstra~\cite{DBLP:journals/mor/Lenstra83}. The running time of his procedure is~$2^{O(n^2)}$ times a polynomial in the encoding length. In fact, Lenstra's algorithm solves the more general \emph{integer programming} problem.   Kannan 
\cite{DBLP:journals/mor/Kannan87} improved this to $n^{O(n)}$ time and polynomial space. It took almost 15 years until Ajtai, Kumar and Sivakumar 
presented a randomized algorithm for $\text{SVP}_2$ with time and space 
$2^{O(n)}$ and a  $2^{O(1+1/\epsilon)n}$ time and space algorithm for \acvp{2} 
\cite{DBLP:conf/stoc/AjtaiKS01,DBLP:conf/coco/AjtaiKS02}. Here \acvp{2} is the problem of finding a lattice vector, whose distance to the target is at most $1+ε$ times the minimal distance. 
Blömer and Naewe~\cite{DBLP:journals/tcs/BlomerN09} extended the randomized sieving algorithm of 
Ajtai et al. to solve $\text{SVP}_p$ and obtain a  $2^{O(n)}$ time and space exact algorithm for \svp[p] and an  $O(1+1/\epsilon)^{2n}$ time algorithm to compute a $(1+ε)$ approximation for \cvp[p]. For \cvp[∞], one has a faster approximation algorithm. Eisenbrand et al.~\cite{DBLP:conf/compgeom/EisenbrandHN11} showed how to boost any constant approximation algorithm for~\cvp[∞]  to a $(1+\epsilon)$-approximation algorithm in time 
$O(\log(1+1/\epsilon))^n$. Recently, this idea was adapted in \cite{cvp_m_and_m} to all $\ell_p$ norms, showing that $(1+\epsilon)$ approximate \cvp[p] can be solved in time $(1+1/\epsilon)^{n/\min(2,p)}$ by boosting the deterministic CVP algorithm for general (even asymmetric) norms with a running time of $(1 + 1/\epsilon)^n$ that was developed by Dadush and Kun~\cite{DBLP:journals/toc/DadushK16}. 

The first deterministic singly-exponential time and space algorithm  for  exact $\text{CVP}_2$ (and $\text{SVP}_2$) was developed by~\cite{DBLP:conf/stoc/MicciancioV10}. The fastest exact algorithms for \svp[2] and \cvp[2]  run in time and space  $2^{n + o(n)}$~\cite{svp_ADRS,gaussian_sampling,DBLP:conf/soda/AggarwalS18}.  Single exponential time and space algorithms for exact $\text{CVP}$ are only known for $\ell_2$.
Whether \cvp[] and the more general integer programming problem can be solved in time $2^{O(n)}$ is a prominent mystery in algorithms.

Recently there has been exciting progress in understanding the \emph{fined grained complexity} of exact and constant approximation algorithms for \cvp[]~\cite{aggarwal2019fine,cvp_hard_seth,svp_hard_seth}.  Under the assumption of the \emph{strong exponential time hypothesis (SETH)} and for  $p≠0 \pmod{2}$,  exact \cvp[p]  cannot be solved in time $2^{(1-\epsilon)d}$. Here $d$ is the \emph{ambient dimension} of the lattice, which is the number of vectors in a basis of the lattice. Under the assumption of a \emph{gap-version} of the  strong exponential time hypothesis \emph{(gap-SETH)}  these lower bounds also hold for the approximate versions of \cvp[p]. More precisely, for each $ε>0$ there exists a constant $γ_ ε>1$ such that there exits no $2^{(1-\epsilon)d}$ algorithm that computes a $γ_ε$-approximation of \cvp[p].

Unfortunately, the currently fastest algorithms for $\cvp[p]$ resp. $\svp[p]$ do not match these lower bounds, even for large approximation factors. These algorithms are based on randomized sieving, \cite{DBLP:conf/stoc/AjtaiKS01,DBLP:conf/coco/AjtaiKS02}. Many lattice vectors are generated that are then, during many stages, subtracted from each other to obtain shorter and shorter vectors w.r.t. $\ell_p$ (resp. any norm) until a short vector is found. However, the algorithm needs to start out with sufficiently many lattice vectors just to guarantee that two of them are close. This issue directly relates to the \emph{kissing number} (w.r.t. some norm) which is the maximum number of unit norm balls that can be arranged so that they touch another given unit norm ball. In the setting of sieving, this is the number of vectors of length $r$ that are needed to guarantee that the difference of two of them is strictly smaller than $r$. Among all known upper bounds on the kissing numbers, the best (i.e. smallest) upper bound is known for $\ell_2$ and equals $2^{0.401 n}$, \cite{kabatiansky1978bounds}. For $\ell_2$ the fastest such approximation algorithms require time $2^{0.802 n}$ - the square of the kissing number w.r.t. $\ell_2$. For $\ell_{\infty}$ the kissing number equals $3^n - 1$ which is also an upper bound on the kissing number for any norm. The current best constant factor approximation algorithms for \svp[\infty] and \cvp[\infty] require time $3^n$, their counterparts w.r.t. $\ell_p$ require even more time, see ~\cite{svp_l_infty,svp_mukh_l_p}. This then suggests the question, originally raised by Aggarwal et al. in ~\cite{aggarwal2019fine} for $\ell_{\infty}$, whether the kissing number w.r.t. $\ell_p$ is a natural lower bound on the running time of \svp[p] resp. \cvp[p]. 

\medskip 
\noindent
Our results indicate otherwise. For constant approximation factors, we are able to reduce these problems w.r.t. $\ell_p$ to another lattice problem but w.r.t. $\ell_2$. This directly improves the running time of the algorithms for $\ell_p$ norms that hinge on the kissing number. Furthermore, given that the development of algorithms for $\ell_2$ has been much more dynamic than for arbitrary $\ell_p$ norms and the difficulty of establishing hardness results for $\ell_2$, there is hope to find still faster algorithms for $\svp[2]$ that may not even rely on the kissing number w.r.t. $\ell_2$. It is likely that this would then improve the situation for $\ell_p$ norms as well.

%
\medskip 
\noindent 
Our \emph{main results} are resumed in the following theorem.

\begin{theorem*}
  For each $ε>0$, there exists a constant $γ_ ε$ such that a $γ_ ε$ approximate solution to  \cvp[p], as well as to \svp[p] for $p \in [1, \infty]$ can be found in time $2^{(0.802 + \epsilon) n}$. 
\end{theorem*}
\noindent 
Our main idea is to use coverings in order to obtain a constant factor approximation to the shortest resp. closest vector w.r.t. $\ell_p$ by using a (approximate) shortest vector algorithm w.r.t. $\ell_2$. We need to distinguish between the cases $p \in [2,\infty]$ and $p \in [1,2)$.
For $p \in [2,\infty]$, we show that exponentially many short vectors w.r.t. $\ell_2$ cannot all have large pairwise distance w.r.t. $\ell_p$. This follows from a bound on the number of $\ell_p$ norm balls scaled by some constant that are required to cover the $\ell_2$ norm ball of radius $n^{1/2 - 1/p}$. The final procedure is then to sieve w.r.t. $\ell_2$ and to pick the smallest non zero pairwise difference w.r.t. $\ell_p$ of the (exponentially many) generated lattice vectors. This yields a constant factor approximation to the shortest resp. closest vector w.r.t. $\ell_p$, $p \in [2, \infty]$. 
For $p \in [1,2)$, we use a more direct covering idea. There is a collection of at most $2^{\epsilon n}$ balls w.r.t. $\ell_2$, whose union contains the $\ell_p$ norm ball but whose union is contained in the $\ell_p$ norm ball scaled by some constant. This leads to a simple algorithm for $\ell_p$ norms ($p \in [1,2)$) by using the approximate closest vector algorithm w.r.t. $\ell_2$ from this paper.

\medskip
\noindent
This paper is organized as follows. In Section \ref{sec:coverings} we present the main idea for $p = \infty$ that also applies to the case $p \geq 2$. 
In Section \ref{sec:appr-textscsvp_-text} we first reintroduce the list-sieve method originally due to \cite{svp_mic_voulg} but with a slightly more general viewpoint, we resume this in Theorem \ref{thr:listred_properties}. We then present in detail our approximate \cvp[\infty] resp. \svp[\infty] algorithm and extend this idea resp. algorithm to $\ell_p$, $p \geq 2$. This is Theorem \ref{thr:1}. Finally, in Section \ref{sec:ell_1}, using the covering technique from Section \ref{sec:coverings} and our approximate \cvp[2] algorithm from Section \ref{sec:list-sieve}, we show how to solve approximate \cvp[p] for $p \in [1,2)$. This is Theorem \ref{thr:2}.

\section{Covering balls with boxes}
\label{sec:coverings}

We now outline the our  main idea in the setting of an approximate $\svp[∞]$ algorithm. 
 Let us assume that the shortest vector of $ℒ$ w.r.t. $ℓ_{∞}$ is $\vs ∈ ℒ \setminus \{0\}$. We can assume that the lattice is scaled such that $\| \vs\|_{∞}=1$ holds.  The euclidean norm of $\vs$ is then bounded by $\sqrt{n}$. Suppose now that there is a procedure that, for some constant $γ >1$ independent of $n$, generates distinct lattice vectors $\vv_1,\dots,\vv_N ∈ ℒ$ of length at most $\|\vv_i\|_2 ≤γ \sqrt{n}$.
\begin{figure}[h]
  \begin{center}
    \includegraphics[width=0.30\textwidth]{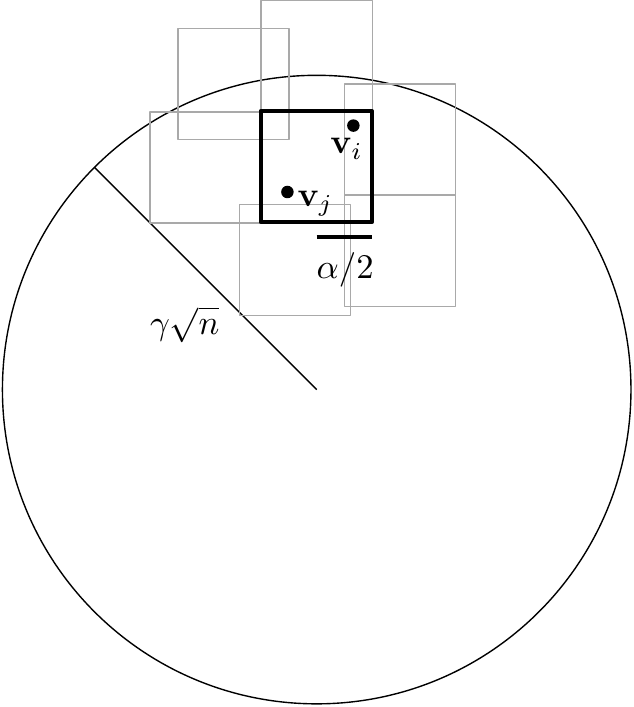}
  \end{center}
  \caption{The difference $\vv_i - \vv_j$ is an $\alpha$-approximate shortest vector w.r.t. $ℓ_∞$.}\label{fig:1}
\end{figure}

How large does the number of vectors $N$ have to be such that we can guarantee   that there exists two indices $i ≠j$ with 
\begin{equation}
  \label{eq:1}
  \|\vv_i - \vv_j \|_∞ ≤ α,
  \end{equation}
  where $α ≥1$ is the approximation guarantee for $\svp[∞]$ that we want to achieve?
  Suppose that $N$ is larger than the minimal number of copies of the box $ (α/2) B_∞^n$ that are required to cover the ball $\sqrt{n}B_2^n$.  Here $B_p^n = \{ x ∈ ℝ^n ： \|x\|_p≤1\}$ denotes the unit ball w.r.t. the $ℓ_p$-norm.
  Then, by the pigeon-hole  principle, two different vectors $\vv_i$ and $\vv_j$  must be in the same box. Their difference satisfies~\eqref{eq:1} and thus is an $α$-approximate shortest vector w.r.t.~$ℓ_∞$, see Figure~\ref{fig:1}.

Thus we are interested in the \emph{translative covering number} $N(\sqrt{n}B_2^n, aB_{\infty}^n)$, which is the number of translated copies of the box $aB_{\infty}^n$ that are needed to cover the $ℓ_2$-ball of radius $\sqrt{n}$. In the setting above, $a$ is the constant $α/(2γ)$. Covering problems like these have received considerable attention in the field of convex geometry, see~\cite{artstein2015weighted,Marton:covering}. 
These techniques rely on the classical \emph{set-cover problem} and the logarithmic integrality gap of its standard LP-relaxation, see, e.g.~\cite{vazirani2013approximation,setcover_chvatal}.
To keep this paper self-contained, we briefly explain how this can be applied to our setting.

If  we cover the finite set 
$({1}/{n})\Z^n \cap \sqrt{n}B_2^n$ with cubes whose centers are on the grid $({1}/{n}) \Z^n$, then by increasing the side-length of those cubes by an additive ${1}/{n}$, one obtains a  full covering of $\sqrt{n}B_2^n$.  Thus we  can focus  on the corresponding \emph{set-covering problem} with ground set ${U} = (1/n) \Z^n \cap \sqrt{n}B_2^n$ and sets
\begin{displaymath}
  S_t = {U}  ∩ a B_∞^n + t, \, t ∈ (1/n) \, ℤ^n, 
\end{displaymath}
ignoring  empty sets. 
An element of the ground set is contained in  exactly $|(1/n) ℤ^n ∩ a B_∞^n|$ many sets.  Therefore, by assigning each
element of the ground set  the fractional value $1/ |(1/n) ℤ^n ∩ a B_∞^n|$, one obtains a feasible fractional covering. The weight of this fractional covering is
\begin{displaymath}
  \frac{T}{ |({1}/{n}) ℤ^n ∩ a B_∞^n|} 
\end{displaymath}
where $T$ is the number of sets.  Clearly, if a cube intersects $\sqrt{n}B_2^n$, then its center  is contained in the \emph{Minkowski sum} $\sqrt{n}B_2^n + a B_∞^n$ and thus the weight of the fractional covering is 
\begin{displaymath}
  \frac{|(\sqrt{n}B_2^n + a B_∞^n) ∩ \frac{1}{n}ℤ^n|}{ |\frac{1}{n} ℤ^n ∩ a B_∞^n|} = O \left(
    \frac{\vol(\sqrt{n}B_2^n + a B_∞^n) }{ \vol(a B_∞^n)}
  \right)
\end{displaymath}
Since the size of the ground-set is bounded by $n^{O(n)}$ and since the integrality gap of the set-cover LP is at most  the logarithm of this size, one obtains 
\begin{alignat}{1}\label{setcover_ineq}
N(\sqrt{n}B_2^n, aB_{\infty}^n) \leq \poly(n) \, \frac{\vol(\sqrt{n}B_2^n + aB_{\infty}^n)}{\vol(aB_{\infty}^n)} 
\end{alignat}
By \emph{Steiner's formula}, see \cite{Gruber_2007,schneider_2013,henk1997basic}, the volume of $K + t B_2^n$ is a polynomial in $t$, with coefficients $V_j(K)$ only depending on the convex body $K$:
$$\vol(K + tB_2^n) = \sum_{j=0}^{n}V_j(K)\vol(B_2^{n-j})t^{n-j}$$
For  $K = aB_{\infty}^n$, $V_j(K)=(2a)^j \binom{n}{j}$. Setting $t = \sqrt{n}$, the resulting expression has been evaluated in \cite[Theorem~7.1]{ball+cube}. 

\begin{theorem}[\cite{ball+cube}] \label{vol:ball+cube}
	Denote by $\binentropy$ the binary entropy function and let $\phi \in (0,1)$ the unique solution to
	\begin{alignat}{1}\label{ineq:unique_solution}
		\frac{1-\phi^2}{\phi^3} = \frac{2a^2}{\pi}
	\end{alignat}
	Then
	\begin{alignat*}{1}
		\vol(a B_{\infty}^n + \sqrt{n}B_2^n) = O(2^{n[\binentropy(\phi) + (1-\phi)\log(2a) + \frac{\phi}{2}\log(\frac{2 \pi e}{\phi})]})
	\end{alignat*}
\end{theorem}
\noindent 
Using this bound in inequality (\ref{setcover_ineq}) and simplifying, we find
$$N(\sqrt{n}B_2^n, aB_{\infty}^n) \leq \poly(n) \,  2^{n[\binentropy(\phi) + \frac{\phi}{2}\log(\frac{2 \pi e}{\phi})]}$$
Both $\binentropy(\phi)$ and $\frac{\phi}{2}\log(\frac{2 \pi e}{\phi})$ decrease to $0$ as $\phi$ decreases to $0$. Since $\phi$, the unique solution to ($\ref{ineq:unique_solution}$), satisfies $\phi \leq \sqrt[3]{(\pi/2)}a^{-\frac{2}{3}}$, we obtain the following bound.

\begin{lemma}
	\label{lem:1}
	For each $ε>0$, there exists $a_ε \in \R_{>0}$ independent of $n$, such that 
	\begin{displaymath}
		N(\sqrt{n}B_2^n, a_εB_{\infty}^n) \leq 2^{ε n}.  
	\end{displaymath}
\end{lemma}
\noindent
Going back to the idea for an approximate \svp[\infty] algorithm, we will use Lemma \ref{lem:1} with $\epsilon = 0.401$. If we generate $2^{0.401 n}$ distinct lattice vectors of euclidean length at most $\gamma \sqrt{n}$, then there must exist a pair of lattice vectors with pairwise distance w.r.t. $\ell_{\infty}$ shorter than $2\gamma a_{0.401}$. We find it by trying out all possible pairwise combinations, this takes time $2^{0.802 n}$.\\
\noindent
The main idea for approximate \svp[p] is similar. Set $\tilde{\vs}$ the shortest vector in $\mathscr{L}$ w.r.t. $\ell_p$ and scale the lattice so that $\norm{\tilde{\vs}}_p = 1$. The euclidean norm of $\tilde{\vs}$ is bounded by $n^{1/2 - 1/p}$. Again, we can consider the question of how many different lattice vectors there have to be within a ball of radius $\gamma n^{1/2 - 1/p}$ so that we can guarantee that there exist two lattice vectors with constant pairwise distance w.r.t. $\ell_p$. This leads us to consider the translative covering number $N(n^{1/2-1/p}B_2^n, aB_p^n)$. Since $n^{-1/p}B_{\infty}^n \subseteq B_p^n$, the following is immediate from Lemma \ref{lem:1}.

\begin{lemma}
	\label{lem:2}
	For each $ε>0$, there exists $a_ε \in \R_{>0}$ independent of $n$, such that 
	\begin{displaymath}
	N(n^{1/2-1/p}B_2^n, a_εB_{p}^n) \leq 2^{ε n}.  
	\end{displaymath}
\end{lemma}

\section{Approximate $\textsc{CVP}_{p}$ for $p \geq 2$}
\label{sec:appr-textscsvp_-text}

We now describe our main contribution. As we mentioned already, $\svp[2]$  can be approximated up to a constant factor in time $2^{(0.802 +ε)   n}$ for each $ε>0$. This follows from a careful analysis of the \emph{list sieve} algorithm of Micciancio and Voulgaris~\cite{svp_mic_voulg}, see~\cite{sieving2LWXZ,sieving2PS}. The running time and space of this algorithm is directly related to the \emph{kissing number} of the $ℓ_2$-norm. The running time is the square of the best known upper bound by Kabatiansky and Levenshtein~\cite{kabatiansky1978bounds}. 

The main insight of our paper is that the current list-sieve variants can be used to approximate $\svp[p]$ and $\cvp[p]$ by testing all pairwise differences of the generated lattice vectors. 

\subsection{List sieve}
\label{sec:list-sieve}

We begin by describing the list-sieve method~\cite{svp_mic_voulg} to a level of detail that is necessary to understand our main result. Our exposition follows closely the one given in~\cite{sieving2PS}. Let $ℒ(B)$ be a given lattice and $\vs ∈ℒ$ be an unknown lattice vector. This unknown lattice vector $\vs$ is typically the shortest, respectively closest vector in $ℒ(B)$. 

The list-sieve algorithm has two stages. The input to the \emph{first stage} of the algorithm is an LLL-reduced lattice basis $B$ of $ℒ(B)$, a constant $ε>0$ and a guess $μ$ on the length of $\vs$ that satisfies
\begin{equation}
  \label{eq:2}
  \|\mathbf{s}\|_2 ≤ μ ≤ (1+ 1/n) \|\mathbf{s}\|_2. 
\end{equation}
The first stage then constructs a list of lattice vectors $L ⊆ℒ(B)$ that is random. This list of lattice vectors is then passed on to the second stage of the algorithm.

The \emph{second stage} of the algorithm proceeds by 
sampling points $\vy_1,\dots,\vy_N$ uniformly and independently at random from the ball
\begin{displaymath}
  (ξ_ε⋅ μ) B_2^n, 
\end{displaymath}
where $ ξ_ε$ is an explicit constant depending on $ε$ only. It then  
 transforms these points via a deterministic algorithm $\listred_L$ into lattice points
\begin{displaymath}
  \listred_L(\vy_1),\dots,\listred_L(\vy_N) ∈ ℒ(B). 
\end{displaymath}
The deterministic algorithm $\listred_L$ uses the list $L⊆ℒ(B)$ from the first stage. 

\begin{figure}[h]
  \begin{center}
    \includegraphics[width=0.40\textwidth]{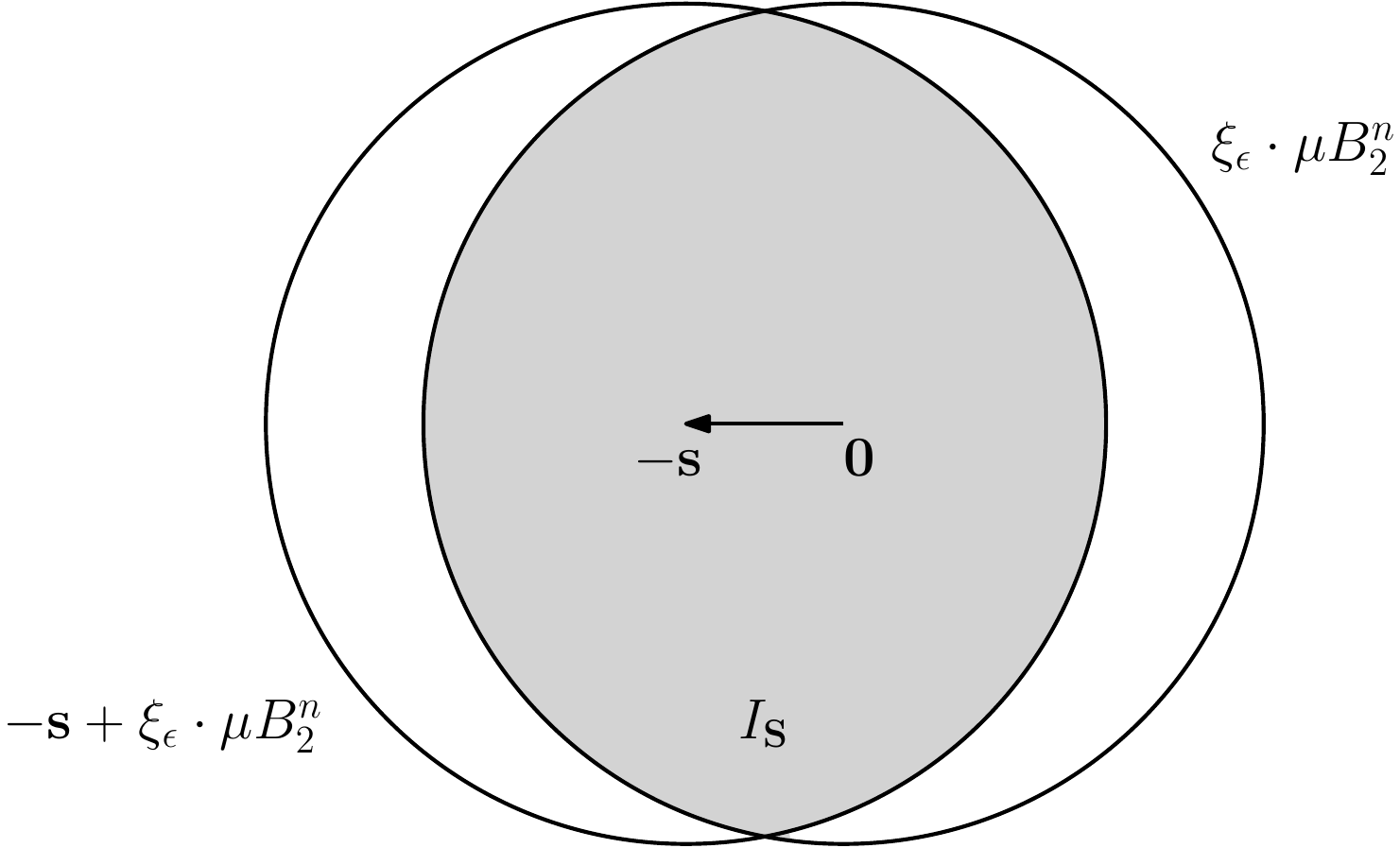}
  \end{center}
  \caption{The lens $I_\vs$} \label{fig:2}
\end{figure}

As we mentioned above, the list $L⊆ℒ(B)$ that is  used by the deterministic algorithm $\listred_L$ is random.  We will show the following theorem in the next section. The novelty compared to the literature is the reasoning about \emph{pairwise differences} lying in \emph{centrally symmetric}  sets.  In this theorem, $ε>0$ is an arbitrary constant, $ξ_ε$ as well as $c_ε$ are explicit constants and $K$ is some centrally symmetric  set. Furthermore, we assume that $μ$ satisfies~\eqref{eq:2}.

The theorem reasons about an area $I_\vs$ that is often referred as the  \emph{lens}, see Figure~\ref{fig:2}. 
The lens was  introduced by Regev as a conceptual modification to facilitate the proof of the original AKS algorithm~\cite{regev_lecturenotes}.   
\begin{equation}
  \label{eq:4}
  I_{\vs} = (\xi_ε ⋅\mu)  B_2^n∩  \left(-\vs\ +  (\xi_ε ⋅\mu) B_2^n  \right) 
\end{equation}

\begin{theorem}\label{thr:listred_properties}
With probability at least $1/2$, the list $L$ that was generated in the first stage  satisfies the following. If  $\,\vy_1, \cdots, \vy_N$ are chosen independently and uniformly at random  within  $B_2^n(0,\xi_{\epsilon}\mu)$ then 
\begin{enumerate}[i)]
  \item \label{item:1}

    The probability of the event that two different samples $ \vy_i, \vy_j$ satisfy 
    \begin{displaymath}      
      \vy_i, \vy_j \in I_{\mathbf{s}} \text{ and } \listred_L(\vy_i) - \listred_L(\vy_j) \in K
  \end{displaymath}
  is at most twice the probability of the event that two different samples  $ \vy_i, \vy_j$ satisfy 
  \begin{displaymath}
  \listred_L(\vy_i) - \listred_L(\vy_j) \in K +\vs 
    \end{displaymath}
  \item \label{item:2}
    For each sample $\vy_i$ the probability of the event
    \begin{displaymath}
    \norm{\listred_L(\mathbf{y}_i)}_2 \leq c_{\epsilon}\norm{\mathbf{s}}_2 \text{ and } \vy_i \in I_{\mathbf{s}}
  \end{displaymath}
  is at least  $2^{-\epsilon n}$. 
  \end{enumerate}  
\end{theorem}
\noindent 
The complete procedure, i.e. the construction of the list $L$ in stage one and applying $\listred_L$ to the $N$ samples $\vy_1,\dots,\vy_n$  in stage two  takes time $N 2^{(0.401 + \epsilon)n} + 2^{(0.802 + \epsilon)n}$ and space $N + 2^{(0.401 + \epsilon)n}$.
\smallskip

The proof of Theorem~\ref{thr:listred_properties} follows verbatim from Pujol and Stehlé~\cite{sieving2PS}, see also \cite{sieving2LWXZ}. In~\cite{sieving2PS},  $\vs$ is a shortest vector w.r.t. $ℓ_2$. But this fact is never used in the proof and in the analysis.   Part~\ref{item:2}) follows from Lemma~5 and Lemma~6 in \cite{sieving2PS}. Their probability of a sample being  in the  lens $I_\vs \subseteq \xi \norm{\vs}_2 B_2^n$ depends only on $\xi$ (corresponding  to our $\xi_{\epsilon}$). By choosing $\xi$ large enough, this happens with probability at least $2^{-εn}$. Their Lemma~6 then guarantees that the list $L$, with probability $1/2$, when $\vy_i \sim I_\vs$ is sampled uniformly, returns a lattice vector of length at most  $r_0 \norm{\vs}_2$ ($r_0$ corresponds to our $c_{\epsilon}$). This corresponds to part ~\ref{item:2}) in our setting. The size of their list (denoted by $N_T$) is bounded above by $2^{(0.401 + \delta)n}$ where $\delta>0$ decreases to $0$ as the ratio $r_0/\xi$ increases, this is their Lemma~4. 

Finally, part~\ref{item:1}) also follows from Pujol and Stehlé~\cite{sieving2PS}. It is in their proof of correctness, Lemma~7, involving the lens $I_\vs$. We briefly comment on our general viewpoint. Given $\mathbf{y} \sim (\xi \cdot \mu) B_2^n$, the algorithm
computes the linear combination w.r.t. to the lattice basis $\vb_1,\dots,\vb_n$ 
\begin{displaymath}
  \vy = ∑_{i=1}^n λ_i \vb_i 
\end{displaymath}
and then the \emph{remainder}
\begin{displaymath}
  \vy \pmod{ℒ} = ∑_{i=1}^n ⌊λ_i⌋ \vb_i.  
\end{displaymath}
The important observation is that this remainder is the same for all vectors $\vy + \vv, \, \vv ∈ ℒ$. Next, it keeps reducing the remainder w.r.t. the list, as long as the length decreases. This results in a vector of the form
\begin{displaymath}
  \vy \pmod{ℒ} - \vv_1 - \cdots - \vv_k, \, \text{ for some } \vv_i ∈ L. 
\end{displaymath}
The output $\listred_L(\vy)$ is then 
\begin{displaymath}
  \vy \pmod{ℒ} - \vv_1 - \cdots - \vv_k + \vy ∈ ℒ. 
\end{displaymath}
The algorithm bases its decisions on $\vy \pmod{ℒ}$ and not on $\vy$ directly. This is why one can imagine that, after $\vy \pmod{ℒ}$ has been created, one applies a bijection $τ$  of the ball $\tau(\cdot): \xi \mu B_2^n \rightarrow \xi \mu B_2^n$ on $\vy$ with probability $1/2$. For  $\vy \in I_\vs$ one has  $τ(\vy) = \vy +  \vs$. We refer to~\cite{sieving2PS} for the definition of $τ$. Since $τ$ is a bijection, the result of  applying $τ(\vy)$ with probability $1/2$  is distributed uniformly. This means that for $\vy ∈I_s$ this modified but equivalent procedure outputs $\listred_L(\vy)$ or $\listred_L(\vy)+\vs$, both with probability $1/2$. If $\listred_L(\vy_i) - \listred_L(\vy_j) ∈ K$, we toss a for $i$ and $j$ each. With probability $1/2$, their difference is in $\pm K + \vs$. 

\subsection{Approximation to $\textsc{CVP}_p$ and $\textsc{SVP}_{p}$ for $p \in [2,\infty]$}
We now combine Theorem~\ref{thr:listred_properties} with the covering ideas presented in Section~\ref{sec:coverings}.
\begin{theorem}
  \label{thr:1}

  For $p \geq 2$, there is a randomized algorithm that  computes with constant probability a 
  constant factor (depending on $\epsilon$) approximation to  $\textsc{CVP}_{p}$ and  $\textsc{SVP}_{p}$ respectively.
The algorithm runs in time  $2^{(0.802 + \epsilon)n}$ and it requires space  $2^{(0.401 + \epsilon)n}$. 
\end{theorem}
\noindent 
In short, the algorithm is the standard list-sieve algorithm with a slight twist: \emph{Check all pairwise differences}. \\
We first present in detail the case $p = \infty$. Even though there is an approximation preserving reduction from \svp[] to \cvp[], \cite{cvp_harderthan_svp},  we present separately the case \svp[] and \cvp[] to highlight the ideas from Section \ref{sec:coverings} and Theorem \ref{thr:listred_properties}. The case $p \geq 2$ then follows from this, we briefly comment on it.

\begin{proof}[Proof for $p = \infty$]
We assume that the list $L$ that was computed in the fist stage satisfies the properties described in Theorem~\ref{thr:listred_properties}. Recall that this is the case with probability at least $1/2$. \\
  We first consider $\textsc{SVP}_{\infty}$.  Choose $a > 0$ such that $ N(\sqrt{n}B_2^n, aB_{\infty}^n) \leq  2^{0.401 n}$ and let  $\vs$ be a shortest vector w.r.t. $\ell_{\infty}$. Furthermore let  $\mu > 0$ such that $\norm{\vs}_{2} \leq \mu < (1 + \frac{1}{n})\norm{\vs}_{2}$ as above. Since $\norm{\vs}_2 \leq \sqrt{n}\norm{\vs}_{\infty}$  we have $N(c_{\epsilon} \norm{\mathbf{s}}_2 B_2^n, c_{\epsilon} a \norm{\mathbf{s}}_{\infty} B_{\infty}^n) \leq 2^{0.401 n}$. This means that, if  $⌈2^{0.401 n}⌉+1$ lattice vectors  are contained in the ball $c_ε \|\mathbf{s}\|_2 B_2^n$ at least two of them have $ℓ_∞$-distance bounded by $2 c_ε a$ which is a constant. 
  
  Set $N = 2 ⋅ ⌈2^{(\epsilon + 0.401) n} +1⌉ $ and  $\{\mathbf{y}_1, \ldots, \mathbf{y}_N\} \stackrel{iid}{\sim} B_2^n(0,\xi_{\epsilon}\mu)$ uniformly and independently at random. By Theorem~\ref{thr:listred_properties}~\ref{item:2})
and by the Chebychev inequality, see~\cite{sieving2PS},  the following event  has probability at least $1/2$. 
  \begin{quote}
    (Event $A$): There is a subset $S ⊆ \{1,\dots,N\}$ with $S =  ⌈ 2^{0.401 n} ⌉+1 $  such that for each $i ∈ S$
  \begin{equation}
    \label{eq:6}
    \vy_i ∈I_\vs \text{ and } \|\listred_L(\vy_i) \|_2 ≤ c_ε \|\vs\|_2. 
  \end{equation}
\end{quote}
This event is the disjoint union of the event $A∩B$ and $A ∩ \overline{B}$,   where $B$ denotes the event where the  vectors $\listred_L(\vy_i),\, y_i ∈I_\vs$ are all distinct. Thus
\begin{displaymath}
   \Pr(A)= \Pr(A ∩B) + \Pr(A ∩ \overline{B}).
\end{displaymath}
The probability of at least one of the events $A ∩B$ and $A ∩ \overline{B}$ is bounded below by $1/4$. In the event $A ∩B$, there exists $i≠j$ such that
\begin{displaymath}
  \|\listred_L(\vv_i) - \listred_L(\vv_j)\|_∞ ≤2 c_ε a.
\end{displaymath}
By Theorem~\ref{thr:listred_properties}~\ref{item:1}) with $K = \{0\}$ one has 
\begin{displaymath}
  \Pr(A ∩ \overline{B})  ≤ 2 \Pr\left(∃i≠j ： \listred_L(\vv_i) - \listred_L(\vv_j) = \vs\right).
\end{displaymath}
Therefore, with constant probability, there exist $i,j ∈ \{1,\dots,N\}$ with
\begin{displaymath}
0<  \| \listred_L(\vy_i)-\listred_L(\vy_j)\|_∞ ≤2 c_ε a. 
\end{displaymath}
We try out all the pairs of $N$ elements, which amounts to $N^2 = 2^{(0.802 + ε')n}$ additional time.


We next describe how list-sieve yields a constant approximation for    $\textsc{CVP}_{\infty}$.  Let $\vw ∈ℒ(B)$ be the closest lattice vector w.r.t. $\ell_{\infty}$ to $\vt ∈ ℝ^n $ and let $\mu > 0$ such that $\norm{\mathbf{t}-\mathbf{w}}_2 \leq \mu < (1 + \frac{1}{n})\norm{\mathbf{t}-\mathbf{w}}_2$. We  use Kannan's embedding technique \cite{DBLP:journals/mor/Kannan87} and define a new lattice $ℒ'$ with basis
\begin{displaymath}
\wt{B} =
\begin{pmatrix}
B & \vt \\
0 & \frac{1}{n}\mu 
\end{pmatrix} ∈ ℚ ^{(n+1)\times (n+1)},
\end{displaymath}
Finding the closest vector to $\vt$ w.r.t. $ℓ_∞$  in $ℒ(B)$  amounts to finding the shortest vector w.r.t.  $ℓ_∞$ in $ℒ'(\wt{B}) ∩\{ \vx ∈ ℝ^{n+1} ： x_{n+1} =  \frac{1}{n}\mu \}$. 
The vector  $\vs = (\vt-\vw,\frac{1}{n}\mu)$ is such a vector and its euclidean length is smaller than $(1 + \frac{1}{n})\mu$. Let $a >0$ be such that
\begin{displaymath}
N(\sqrt{n} B_2^{n}, a B_∞^{n}) ≤2^{0.401 n}. 
\end{displaymath}
This means that there is a covering of the $n$-dimensional ball $ (c_ ε \|\vs\|_2) B_2^{n+1}∩\{ \vx ∈ ℝ^{n+1} ： x_{n+1} =  0 \}$ by  $2^{0.401 n}$ translated copies of $K$, where
\begin{equation}
\label{cvx_body}
K = (c_ε ⋅a (1 + 1/n) \|\vs\|_∞) B_∞^{n+1} ∩  \{\vx ∈ℝ^{n+1} ： x_{n+1} = 0\}. 
\end{equation}
(The factor $(1 + 1/n)$ is a reminiscent of the embedding trick, $\vs$ is $n+1$ dimensional.)
Similarly, we may cover $ (c_ ε \|\vs\|_2) B_2^{n+1}∩\{ \vx ∈ ℝ^{n+1} ： x_{n+1} =  k \cdot \frac{\mu}{n} \}$ for all $k \in \Z$ (such that the intersection is not empty) by translates of $K$. There are only $2c_{\epsilon}(n+1)  + 1$ such layers to consider and so $(2c_{\epsilon}(n+1)  + 1)2^{0.401 n}$ translates of $K$ suffice. The last component of a lattice vector of $ℒ'$ is of the form $k\cdot \frac{\mu}{n}$ and it follows that these translates of $K$ cover all lattice vectors of euclidean norm smaller than $c_{\epsilon}\norm{\vs}_2$, see Figure~\ref{fig:cvp_list}.

\begin{figure}[h]
	\begin{center}
		\includegraphics[width=0.60\textwidth]{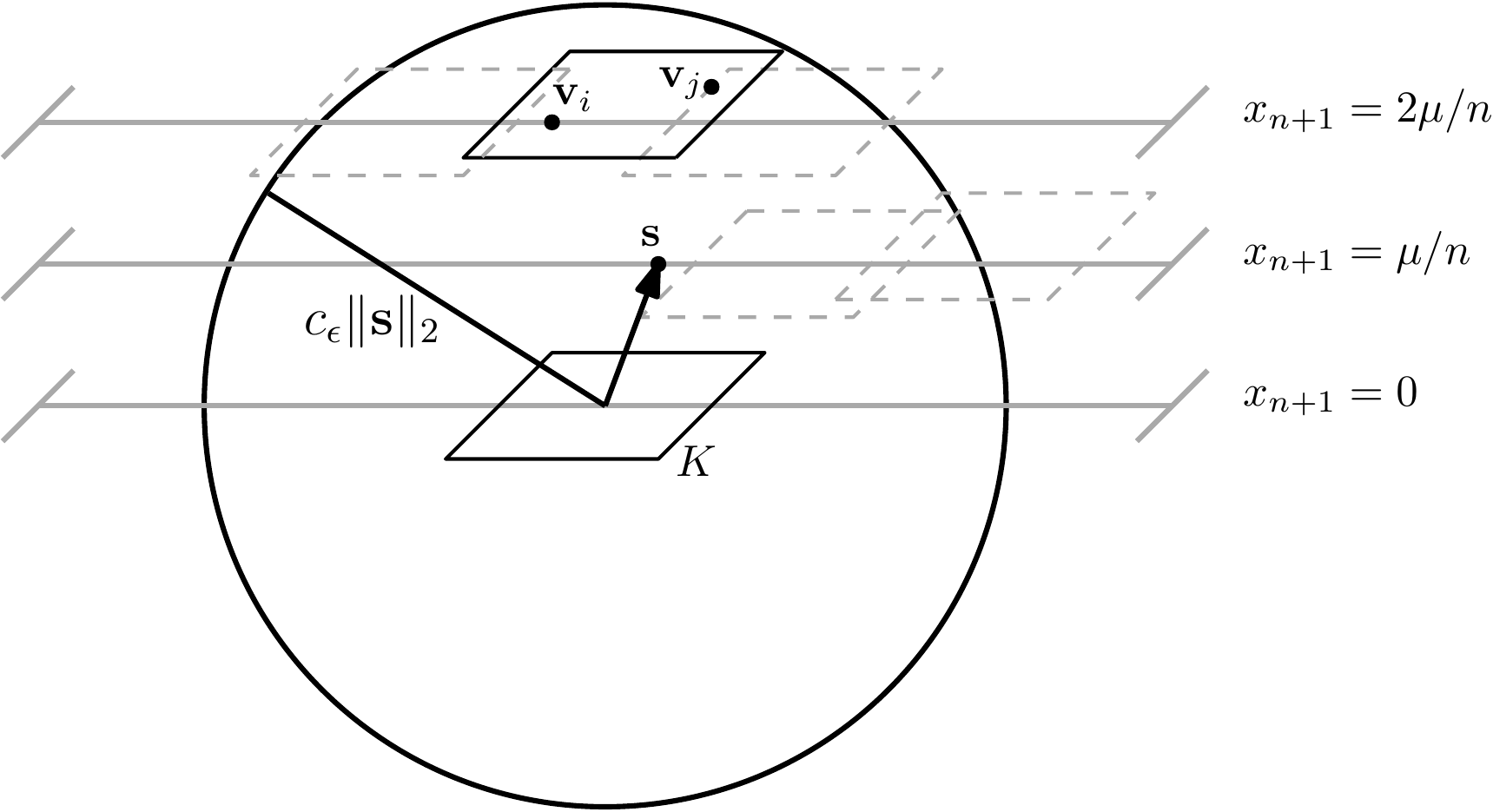}
	\end{center}
	\caption{Covering the lattice points with translates of $K$} \label{fig:cvp_list}. 
\end{figure}

\noindent 
Set $N = ⌈(2c_{\epsilon}(n+1)  + 2) 2^{(\epsilon + 0.401) n}⌉ $ and  sample  again  $\{\mathbf{y}_1, \ldots, \mathbf{y}_N\} \stackrel{iid}{\sim} B_2^n(0,\xi_{\epsilon}\mu)$ uniformly and independently at random. By Theorem~\ref{thr:listred_properties}~\ref{item:2})
and by the Chebychev inequality, see~\cite{sieving2PS},  the following event  has a probability at least $1/2$. 
\begin{quote}
	(Event $A'$): There is a subset $S ⊆ \{1,\dots,N\}$ with $S = (2c_{\epsilon}(n+1)  + 1)2^{0.401 n} + 1 $  such that for each $i ∈ S$
	\begin{equation}
	\label{eq:6}
	\vy_i ∈I_\vs \text{ and } \|\listred_L(\vy_i) \|_2 ≤ c_ε \|\vs\|_2. 
	\end{equation}
\end{quote}
In this case, there exists a translate of $K$ that holds at least two vectors $\listred_L(\vy_i)$ and $\listred_L(\vy_j)$ for different samples $\vy_i$ and $\vy_j$, see Figure~\ref{fig:cvp_list} with $\vv_i, \vv_j \in ℒ'$ instead. Thus, with probability at least $1/2$, there are $i, j \in [N]$ with $\vy_i, \vy_j \in I_{\vs}$ such that
\begin{displaymath}
\listred_L(\vy_i) - \listred_L(\vy_j) \in 2K
\end{displaymath}
Theorem~\ref{thr:listred_properties}~\ref{item:1}) implies that, with probability at least $1/4$, there exist different samples  $\vy_i$ and $\vy_j$ such that
\begin{displaymath}
\listred_L(\vy_i) - \listred_L(\vy_j) \in 2K + \vs 
\end{displaymath}
In this case, the first $n$ coordinates of $\listred_L(\vy_i) - \listred_L(\vy_j)$ can be written of the form $\vt - \vv$ for $\vv \in ℒ$ and the first $n$ coordinates on the right hand side are of the of the form $(\vt - \vw) + \mathbf{z}$, where $\mathbf{z} \in ℒ'$ and $\norm{\mathbf{z}}_{\infty} \leq 2 c_{\epsilon} (1 + 1/n) a  \norm{\vs}_{\infty} = 2 c_{\epsilon} (1 + 1/n) a \norm{\vt- \vw}_{\infty}$. 
In particular, the lattice vector $\vv \in ℒ$ is a $2 a c_{\epsilon}(1 + 1/n) + 1$ approximation to the closest vector to $\vt$.
Since we need to try out all pairs of the $N$ elements, this takes time $N^2 = 2^{(0.802 + \epsilon')n}$ and space $N$.
\noindent
\end{proof}

\begin{remark}
For clarity we have not optimized the approximation factor. There
are various ways to do so.  We remark that for 
$\svp[\infty]$  we actually get a  smaller
approximation factor than the one that we describe.  Let $\tilde{a}$ be such that
$N(\sqrt{n}B_2^n, \tilde{a}B_{\infty}^n) \leq 2^{0.802 n}$, the
algorithm described above yields a $2 c_{\epsilon} \tilde{a}$
approximation instead of a $2 c_{\epsilon} a$ approximation to the
shortest vector. This follows by applying  the \emph{birthday paradox} in the way that it was used by Pujol and Stehlé~ \cite{sieving2PS}. The same
argument also applies to 
$\cvp[\infty]$. Finally, we remark that in the case of \svp[] we have not really used property~\ref{item:1}) of Theorem \ref{thr:listred_properties}. We only use this property to ensure that the generated vectors are different. It is plausible that this can be done more efficiently or with a better approximation factor.
\end{remark}

\begin{proof}[Proof continued, $p \geq 2$]
For \svp[p], $p \geq 2$, we define $\vs$ to be shortest vector w.r.t. $\ell_p$ instead. Since $\norm{\vs}_2 \leq n^{1/2 - 1/p} \norm{\vs}_p$, we simply use Lemma \ref{lem:2} instead of Lemma \ref{lem:1} to conclude that there is some $a > 0$ such that if we have a set of $2^{0.401 n}$ different lattice vectors of (euclidean) length smaller than $c_{\epsilon}\norm{\vs}_2$, then two of them must have pairwise distance smaller than $2c_{\epsilon}a$ w.r.t. $\ell_p$.\\
For \cvp[p], we define $\vw$ to be the closest lattice vector to $\vt$ w.r.t. $\ell_p$. Both $\vs$ and $\mathscr{L}'$ are defined analogously. We will need to replace the convex body $K$ in (\ref{cvx_body}) by 
\begin{displaymath}
K = (c_ε ⋅a (1 + 1/n) \|\vs\|_p) B_p^{n+1} ∩  \{\vx ∈ℝ^{n+1} ： x_{n+1} = 0\}. 
\end{displaymath}
The respective algorithms for \svp[p] and \cvp[p] and the proof of correctness now follow from the case $p = \infty$. In particular, we can use the same parameters $c_{\epsilon}$ and $a$.\\ 
\newline
\noindent
For the important case $p = 2$ we note that we can chose $a = 1$. This yields a approximation to the closest vector with the approximation guarantee $c_{\epsilon}$ matching that of the fastest approximate shortest vector problem w.r.t. $\ell_2$, see \cite{sieving2LWXZ}.

\end{proof}

\section{Approximate $\textsc{CVP}_{p}$ for $p \in [1,2)$}\label{sec:ell_1}

In the previous section, we have extended the approximate \svp[2] solver to yield constant factor approximations to \svp[p] and \cvp[p] for $p \in [2, \infty]$ in time $2^{(0.802 + \epsilon)n}$. From simple volumetric considerations, the technique from the previous section cannot be adapted to solve \svp[p] and \cvp[p] for $p \in [1,2)$ (in single exponential time). Instead, we can use a simple covering technique similar to the one considered by Eisenbrand et al. in \cite{DBLP:conf/compgeom/EisenbrandHN11}. We first show that for any constant $\epsilon > 0$, there is a constant $a_{\epsilon} > 0$, so that the crosspolytope $B_1^n$ can be covered by $2^{\epsilon n}$ balls (w.r.t. $\ell_2$) with radius $(a_{\epsilon}/\sqrt{n})$ and whose union is contained inside the crosspolytope scaled by $a_{\epsilon}$. A similar covering also exists for $B_p^n$. Using the centers of these balls as targets, we can use the approximate \cvp[2] algorithm to solve approximate \cvp[1] resp. \cvp[p]. 
To achieve this, we rely on the set-covering idea and volume computations as outlined in Section \ref{sec:coverings}. The following analogue to Lemma \ref{lem:1} is shown in the appendix.

\begin{figure}[h]
	\begin{center}
		\includegraphics[width=0.50\textwidth]{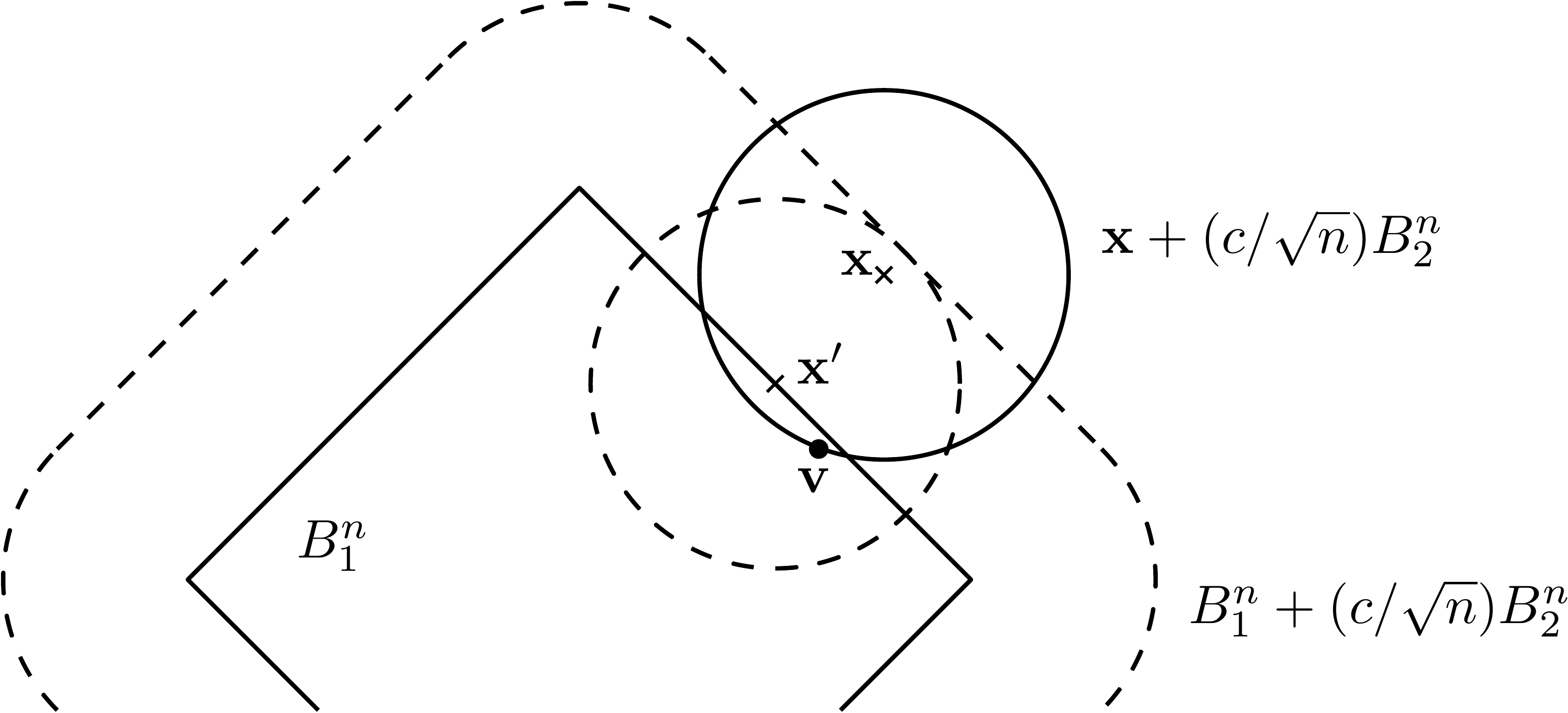}
	\end{center}
	\caption{Generating a covering of $B_1^n$ by $(c/\sqrt{n})B_2^n$} \label{fig:covering_l1}. 
\end{figure}

\begin{lemma}
	\label{lem:1-2}
	For each $ε>0$, there exists $a_ε \in \R_{>0}$ independent of $n$ such that 
	\begin{displaymath}
	\frac{\vol(B_1^n + (a_ε/\sqrt{n})B_2^n)}{\vol((a_ε/\sqrt{n})B_2^n)} \leq 2^{ε n}.  
	\end{displaymath}
\end{lemma}
\noindent
We now sketch the covering procedure for \cvp[1] and \svp[1]. Up to scaling the lattice and a guess on the distance of the closest (resp. shortest) lattice vector $\vv$ to the target $\vt$, we may assume that $1-1/n \leq \norm{\vv- \vt}_1 \leq 1$ (resp. $1- 1/n \leq \norm{\vv}_1 \leq 1$).
We uniformly sample a point $\vx$, \cite{DBLP:journals/jacm/DyerFK91}, within $\vt + B_1^n + (a_ε/\sqrt{n})B_2^n$ (set $\vt = 0$ for \svp[1]) and place a ball of radius $ a_ε/\sqrt{n}$ around $\vx$ (or $\vx'$, the closest point to $\vx$ in $B_1^n$, see Fig.~\ref{fig:covering_l1}). 

By Lemma \ref{lem:1-2}, with probability at least $2^{-\epsilon n}$, $\vv$ is covered by $\vx + (a_{\epsilon}/\sqrt{n})B_2^n$. Running the $c$-approximate (randomized) \cvp[2] algorithm with target $\vx$ (provided $\norm{\vv - \vx}_2 \leq (a_{\epsilon}/\sqrt{n})$), a lattice vector $\vw \in \vx + (c \cdot a_ε/\sqrt{n})B_2^n \subseteq \vt + c \cdot (a_{\epsilon} + 1)B_1^n $ is returned. The lattice vector $\vw$ is thus a $c \cdot (a_{\epsilon} + 1)$ approximation to the closest (resp. shortest) vector. In general, we run the $c$-approximate $\cvp[2]$ algorithm $O(\poly(n)2^{\epsilon n})$ times with targets uniformly chosen within $\vt + B_1^n + (a_ε/\sqrt{n})B_2^n$ and only output the closest of the resulting lattice vectors if it is within $c \cdot (a_{\epsilon} + 1)B_1^n$. This ensures that, if there is lattice vector $\vt$ in $\vt + B_1^n$, a constant factor approximation to $\norm{\vt-\vv}_1$ is found with high probability. \\
The same covering technique can be applied to $B_p^n$, $p \in (1,2)$. By Hölder's inequality,
\begin{displaymath}
B_p^n \subseteq n^{1-1/p}B_1^n \text{ and } n^{1/2-1/p}B_2^n \subseteq B_p^n.
\end{displaymath}
The first of these inclusions implies that for any $\epsilon > 0$, we can pick the same constant $a_{\epsilon}$ as in Lemma \ref{lem:1-2} and cover $B_p^n$ by at most $2^{\epsilon n}$ translates of $a_\epsilon n^{1/2 - 1/p} B_2^n$. 
\begin{displaymath}
\frac{\vol(B_p^n + c n^{1/2 - 1/p}B_2^n)}{\vol(c n^{1/2 - 1/p}B_2^n)} \leq \frac{\vol(n^{1-1/p}B_1^n + c n^{1/2 - 1/p}B_2^n)}{\vol(c n^{1/2 - 1/p}B_2^n)} = \frac{\vol(B_1^n + (c/\sqrt{n})B_2^n)}{\vol((c/\sqrt{n})B_2^n)} 
\end{displaymath}
The second inclusion implies that these translates do not overlap $B_p^n$ by more then a constant factor.
It is then straightforward to adapt the boosting procedure described for \cvp[1] to \cvp[p]. Using the approximate \cvp[2] algorithm from the previous section then implies the following algorithm.

\begin{theorem}
	\label{thr:2}
	There is a randomized algorithm that  computes with constant probability a 
	constant (depending on $\epsilon$) factor approximation to $\textsc{CVP}_{p}$, $p \in [1,2)$.
	The algorithm runs in time  $2^{(0.802 + \epsilon)n}$ and requires space $2^{(0.401 + \epsilon)n}$. 
\end{theorem}
\appendix

\section{Proof of Lemma \ref{lem:1-2}}
Recall that the volume of $K + t B_2^n$ is a polynomial in $t$, with coefficients $V_j(K)$ that only depend on the convex body $K$:
\begin{displaymath}
\vol(K + tB_2^n) = \sum_{j=0}^{n}V_j(K)\vol(B_2^{n-j})t^{n-j}
\end{displaymath}
The coefficients $V_j(K)$ are known as the intrinsic volumes of $K$. The intrinsic volumes of the crosspolytope $B_1^n$ were computed by Betke and Henk in \cite{intrinsic_vol_B_1^n}, and are given by the following formulae:
\begin{displaymath}
V_n(B_1^n) = \frac{2^n}{n!}
\end{displaymath}
and for $0 \leq j \leq n-1$
\begin{displaymath}
V_j(B_1^n) = 2^n\,{n \choose j+1}\frac{\sqrt{j+1}}{j! \sqrt{\pi}^{n-j}} \, \cdot \int_0^{\infty} e^{-x^2}{\left(\int_0^{x/\sqrt{j+1}}e^{-y^2} \,dy\right)}^{n-j-1}\,dx
\end{displaymath}

Given that the upper bound of Lemma \ref{lem:1-2} is exponential in $n$, we do not care about polynomial factors in $n$. For the sake of brevity, we will hide these polynomial factors by "$\lesssim$", i.e. $\poly(n) \lesssim 1$. This already simplifies the intrinsic volumes and, for $1 \leq j \leq n$:
\begin{displaymath}
V_j(B_1^n) \lesssim \frac{2^j}{j!}\, {n \choose j}
\end{displaymath}
The volume of the $k-$dimensional ball $B_2^k$ is given by
\begin{displaymath}
\vol(B_2^k) = \frac{\pi^{k/2}}{\Gamma(k/2 + 1)}
\end{displaymath}
$\Gamma(\cdot)$ is the Gamma function. For $n \in \N$, we have $\Gamma(n+1) = n!$. By Stirling's formula we have the following estimate on $\Gamma(\cdot)$. 
\begin{displaymath}
	\left(\frac{z}{e}\right)^z \lesssim \,\Gamma(z + 1) \lesssim \left(\frac{z}{e}\right)^z
\end{displaymath}
With these estimates at hand, we can now prove Lemma \ref{lem:1-2}.
\begin{alignat*}{1}
\frac{\vol(B_1^n + (c/\sqrt{n})B_2^n)}{\vol((c/\sqrt{n})B_2^n)} &\stackrel{\phantom{(j = \phi \, n)}}{=}  \frac{\sum_{j=0}^{n}V_j(B_1^n)\vol(B_2^{n-j})(c/\sqrt{n)})^{n-j}}{(c/\sqrt{n})^n \vol(B_2^n)}\\
&\stackrel{\phantom{(j = \phi \, n)}}{\lesssim} \sum_{j=0}^{n}\frac{2^j \, n!}{j!(n-j)! j!}\,\frac{n^{j/2}}{c^j} \, \frac{\vol(B_2^{n-j})}{\vol(B_2^n)}\\
&\stackrel{\phantom{(j = \phi \, n)}}{\lesssim} \sum_{j=0}^{n}\frac{(2e)^j \, n^n}{j^j\,(n-j)^{n-j} j^j}\,\frac{n^{j/2}}{c^j} \, \frac{n^{n/2}}{(n-j)^{(n-j)/2}(2\pi e)^{j/2}}\\
&\stackrel{\phantom{(j = \phi \, n)}}{\lesssim} \sum_{j=0}^{n}\frac{n^{3n/2} n^{j/2}}{j^{2j}(n-j)^{3(n-j)/2}} \, \left(\frac{2e}{\pi c^2}\right)^{j/2}\\
&\stackrel{(j = \phi \, n)}{\lesssim} \max_{\phi \in [0,1]}\frac{e^{(3/2)\ln(n)n + \ln(n)n \phi /2}}{e^{2\ln(\phi n)\phi n + (3/2)\ln((1-\phi)n)(1-\phi)n}}\, \left(\frac{2e}{\pi c^2}\right)^{\phi n/2}\\
&\stackrel{\phantom{(j = \phi \, n)}}{\lesssim} \max_{\phi \in [0,1]}e^{-2\ln(\phi)\phi n - 2\ln(1-\phi)(1-\phi)n} \, \left(\frac{2e}{\pi c^2}\right)^{\phi n/2}\\
&\stackrel{\phantom{(j = \phi \, n)}}{=} \max_{\phi \in [0,1]} 2^{2\binentropy(\phi)n}\, \left(\frac{2e}{\pi c^2}\right)^{\phi n/2}
\end{alignat*}
In passing to the second last line, we have added the factor $e^{-(1/2)\ln(1-\phi)(1-\phi)n}$ which is always greater than $1$ for $\phi \in [0,1]$.
$\binentropy(\cdot)$ is the binary entropy function, i.e. $H(\phi) = -\ln(\phi)\phi - \ln(1-\phi)(1-\phi)$. 
$\binentropy(\phi) \leq 1$ for $\phi \in [0,1]$ and $\binentropy(\phi) = \binentropy(1-\phi) \rightarrow 0$ monotonically as $\phi \rightarrow 0$. Thus, for some fixed $c$, the above expression reaches a maximum for some $\phi \in (0,1)$. If we increase $c$, we see that the $\phi^*$ realizing the maximum will decrease which then implies the lemma. This can be shown formally by fixing some $c$ and taking a derivative w.r.t. $\phi$. This will then show that the maximum is reached when $\phi^* = \Theta(\frac{1}{\sqrt{c}})$. \\
Thus, for any $\epsilon > 0$, we can chose $c$ large enough so that Lemma \ref{lem:1-2} holds.

\printbibliography

\end{document}